\documentclass[11pt, letterpaper]{article}
\usepackage[utf8]{inputenc}
\usepackage[margin=1.5cm]{geometry}
\usepackage{titlesec}
\usepackage{tabu}
\usepackage{longtable}
\usepackage{enumitem}
\usepackage{amssymb}
\usepackage{xcolor}
\newlist{selectlist}{itemize}{2}
\setlist[selectlist]{label=$\square$,leftmargin=*,noitemsep,topsep=0pt}

\usepackage{lmodern}
\usepackage{subfigure}
\usepackage{hyperref}
\hypersetup{
    colorlinks=true,
    linkcolor=blue,
    filecolor=magenta,      
    urlcolor=blue,
}

\newcommand{\kreis}[1]{\unitlength1ex\begin{picture}(2.5,2.5)%
\put(0.75,0.75){\circle{2.5}}\put(0.75,0.75){\makebox(0,0){#1}}\end{picture}}

\usepackage{tikz}
\usepackage{amsmath}

\usepackage{multirow}
\usepackage{color, colortbl}
 
\titleformat{\section}[block]{\hspace{1em}\bfseries}{\thesection.}{0.5em}{} 
\titleformat{\subsection}[block]{\hspace{1em}}{\thesubsection}{0.5em}{}

\begin{document}
\begin{flushleft}

\textbf{Article information}\\
\vskip 0.5cm
\textbf{Dataset of a parameterized U-bend flow for Deep Learning Applications}\\
\vskip 0.5cm
\textbf{Authors}\\ Jens Decke*$^1$, Olaf Wünsch$^2$, Bernhard Sick$^1$
\vskip 0.5cm
\textbf{Affiliations}\\ 
Intelligent Embedded Systems, University of Kassel, Wilhelmshöher Allee 73, D-34121 Kassel$^1$ \\
Fluid Dynamics, University of Kassel, Mönchebergstraße 7, D-34125 Kassel$^2$ \\
\vskip 0.5cm
\textbf{Corresponding author’s email address and Twitter handle}\\ jens.decke@uni-kassel.de 
\vskip 0.5cm
\textbf{Abstract}\\ This dataset contains 10,000 fluid flow and heat transfer simulations in U-bend shapes. Each of them is described by 28 design parameters, which are processed with the help of Computational Fluid Dynamics methods. The dataset provides a comprehensive benchmark for investigating various problems and methods from the field of design optimization. For these investigations supervised, semi-supervised and unsupervised deep learning approaches can be employed. One unique feature of this dataset is that each shape can be represented by three distinct data types including design parameter and objective combinations, five different resolutions of 2D images from the geometry and the solution variables of the numerical simulation, as well as a representation using the cell values of the numerical mesh. This third representation enables considering the specific data structure of numerical simulations for deep learning approaches. The source code and the container used to generate the data are published as part of this work. 
\vskip 0.5cm
\textbf{Specifications table}\\
%
%
%
\begin{longtable}{|p{33mm}|p{124mm}|}
\hline
\textbf{Subject}                & Computational Mechanics
\vskip 0.1cm 
\\
\hline                         
\textbf{Specific subject area}  & Numerical modeling of fluid flow and heat transfer phenomena (conjugate heat transfer) using Computational Fluid Dynamics (CFD)
\vskip 0.1cm 
\\
\hline
\textbf{Type of data}           & 
                         Table\newline
                         Image\newline
                         Code files (Python, Bash)\newline
                        
\vskip 0.1cm 
\\
\hline
\textbf{How the data were acquired} & 
Using Python, OpenFOAM and other open-source software. The data was generated on the department's own computing cluster. The code used is publicly available with the dataset.
\vskip 0.1cm 
\\
\hline                         
\textbf{Data format}            & 
                         Raw\newline
                         Filtered
\vskip 0.1cm 
\\                                                    
\hline
\textbf{Description of          
data collection}             & The dataset consists of 10,000 samples of U-bend shapes. Each is described by 28 parameters. These samples have been generated, evaluated and processed using Python and OpenFOAM. Each sample is independently and identically distributed. The performance of each shape is determined by its pressure loss and cooling capacity.
\vskip 0.1cm 
\\                         
\hline                         
\textbf{Data source location}   & 
$\bullet$ Institution: University of Kassel, Department for Intelligent Embedded Systems \newline
                        $\bullet$ City: Kassel \newline
                        $\bullet$ Country: Germany 
\vskip 0.1cm                        
\\
\hline                         
\hypertarget{target1}
{\textbf{Data accessibility}}   & 
         		     Repository name: DaKS – Datenrepository der Universität Kassel \newline
                         Data identification number: 10.48662/daks-17 \newline
                         Direct URL to data: \url{https://daks.uni-kassel.de/handle/123456789/50} \newline  
                         Code identification number: 10.5281/zenodo.7717020 \newline
                         Direct URL to code: \url{https://zenodo.org/badge/latestdoi/612247257}\newline
                         Item included at the time of submission\\      
\hline
{\textbf{Related research article}}   & Decke, J., Schmeißing, J., Botache, D., Bieshaar, M., Sick, B., Gruhl, C. (2022). NDNET: A Unified Framework for Anomaly and Novelty Detection. In: Schulz, M., Trinitis, C., Papadopoulou, N., Pionteck, T. (eds) Architecture of Computing Systems. ARCS 2022. Lecture Notes in Computer Science, vol 13642. Springer, Cham. \newline 
\url{https://doi.org/10.1007/978-3-031-21867-5_13} \newline \\      
\hline  
\end{longtable}


\textbf{Value of the Data}\\
\begin{itemize}
\itemsep=0pt
\parsep=0pt
\item[$\bullet$] Despite its high complexity, the present multiphysics problem can be both interpreted and comprehended from the perspective of fluid dynamics. 
\item[$\bullet$]The dataset is of interest to researchers who are working on the optimization of topologies, shapes or the design of components in general. 
\item[$\bullet$]The data can be used to develop novel algorithms for design optimization. Especially methods from Deep Learning need a large amount of data, which is not yet publicly available. 
\item[$\bullet$] Due to the number of partial differential equations, there are large numbers of target variables, which are particularly interesting for concepts from transfer learning. For instance, it might be possible to reduce the number of partial differential equations that actually have to be solved. 
\item[$\bullet$]Each sample of the dataset is represented in three different ways (parameters, images, cells).
\end{itemize}
\vskip0.0cm
\textbf{Objective}\\
The dataset may be of interest to researchers combining the fields of deep learning and design optimization using numerical simulations. In that area, this dataset can serve as a benchmark, as it is very versatile since each sample is represented as three different data types. Research in multimodal features, transfer learning, active learning and learning from image and/or graph data using supervised, unsupervised and or semi-supervised learning methods are feasible. The dataset was originally used as a case study to demonstrate the versatile applicability of a developed framework for anomaly and novelty detection \cite{NDNET}. To enable other researchers the use this dataset, and for subsequent investigations, this dataset is provided with public accessibility licensed by CC-BY-NC-4.0.
\vskip0.3cm
\textbf{Data Description}\\
The data contains the results of fluid flow and heat transfer simulations to optimize the design of U-bends. The structure of the introduced dataset is shown in Figure~\ref{fig:dataset} and explained below. 

\begin{figure}[ht!]
	\centering
    \includegraphics[width=0.48\textwidth]{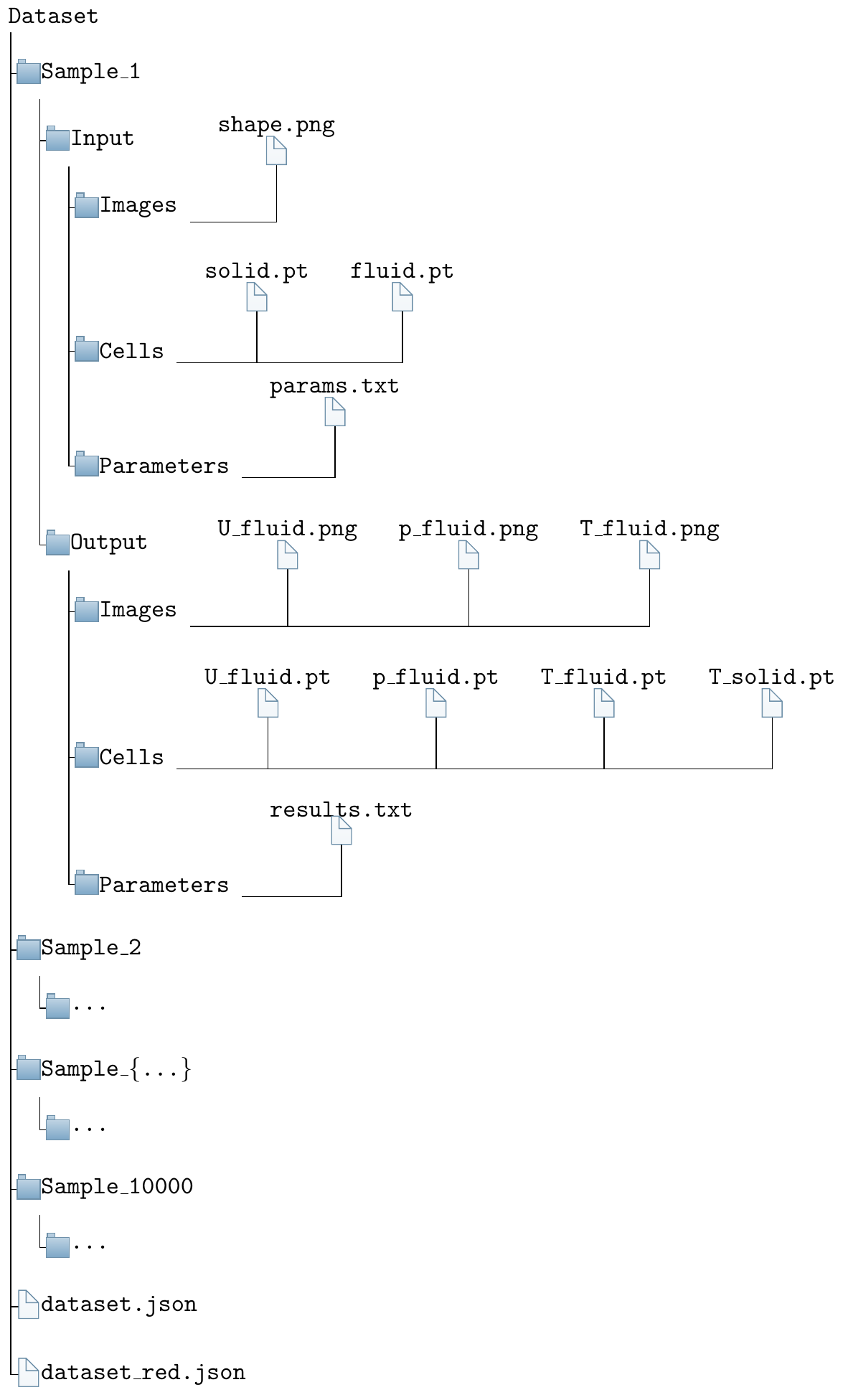}
	\caption{The structure of the present dataset. Each of the i.i.d. samples has its own folder with input and output directories in which the respective files for the three different representations are stored. Structure information of the dataset are provided in the \textit{dataset.json} and \textit{dataset\_red.json}.}
	\label{fig:dataset}
\end{figure}
The dataset includes 10,000 designs that are independent and identically distributed (i.i.d.). Each design has a dedicated folder in the main path and contains an input and an output folder. In these folders are additional directories that contain the data of the different design data formats. These three data formats are parameter values, which define the design, as well as images of the design and the numerical mesh. In the \textit{Parameters} folder the representation of the parameter values are stored. A file \textit{params.txt} is contained in the \textit{Input} representing a vector with 28 design parameters, partly between $-1$ and $1$ as well as partly between $0$ and $1$. The \textit{Output} includes the \textit{results.txt} file with a vector of two target values which are on the one hand the pressure loss between inlet and outlet in $[Pa]$ and on the other hand a performance number for the cooling performance of the design in $[K^{2}m^{2}]$. A second data representation is made in the format of .png images. The input contains a file \textit{shape.png} which represents the fluid as well as the solid area. One image each for the solution variables of the differential equations is included in the output. These solution variables are the velocity vector \textit{U}, the pressure \textit{p} and the temperature \textit{T}. The files are named accordingly (\textit{U\_fluid.png}, \textit{p\_fluid.png}, \textit{T\_fluid.png}). All images are provided in five different resolutions. For the sake of brevity, not all are depicted in Figure~\ref{fig:dataset}. A third way to represent a design is by using the numerical meshes, that had to be generated to compute the designs. This data is located in the \textit{Cells} folder. They are saved in the format of PyTorch \cite{Pytorch} tensors. The files \textit{solid.pt} and \textit{fluid.pt} are provided for the input. These contain information about the coordinates of the respective cell centers of the numerical mesh. The output contains a file for each solution variable of the differential equations representing the solutions for the cell centers (\textit{U\_fluid.pt}, \textit{p\_fluid.pt}, \textit{T\_fluid.pt}, \textit{T\_solid.pt}). In addition, there is a \textit{dataset.json} file in the main path of the dataset, which contains the data structure of all folders and files of the dataset. A  file \textit{dataset\_red.json} is stored in which the structural information about all successfully simulated samples is stored. Depending on the desired investigations, it may be advantageous to use one or the other \textit{json} file.

The functionality of the source code that has generated the data is described in detail in the Experimental Design section. The file \textit{main.py} is the primary file, which starts the generation of the data. With the file \textit{export\_batch\_file.py} a new process and with it the generation of one sample is triggered. The file \textit{export\_blockmesh\_file.py} is the interface between Python and OpenFOAM. The \textit{read\_folder.py} is a utility function with which the \textit{*.json} files are generated. The file \textit{Allrun\_template.sh} is used to pass commands to the podman containers.

\vskip0.3cm
\textbf{\large{Experimental design, materials and methods}}\\
In this section, the parameterization of the design space, the material properties, the boundary conditions and two different objective functions are first presented, and finally the experimental setup is explained.\\
\vskip0.1cm
 \textbf{Parameterization of the design space:} The parameterized model developed in this work is based on a benchmark test case from the \textit{von Karman Institute for Fluid Dynamics}. It was originally introduced by Verstraete et al. \cite{Optimization11Verstraete}. The original model was extended with more design parameters and a solid region which was introduced by Goeke~and~Wünsch \cite{goeke2}. The original three-dimensional problem is simplified to two dimensions as shown in Figure~\ref{fig:DesignSpaceParameter}. 

 \begin{figure}[ht]
	\centering
    \scalebox{1}{\includegraphics[width=.58\textwidth]{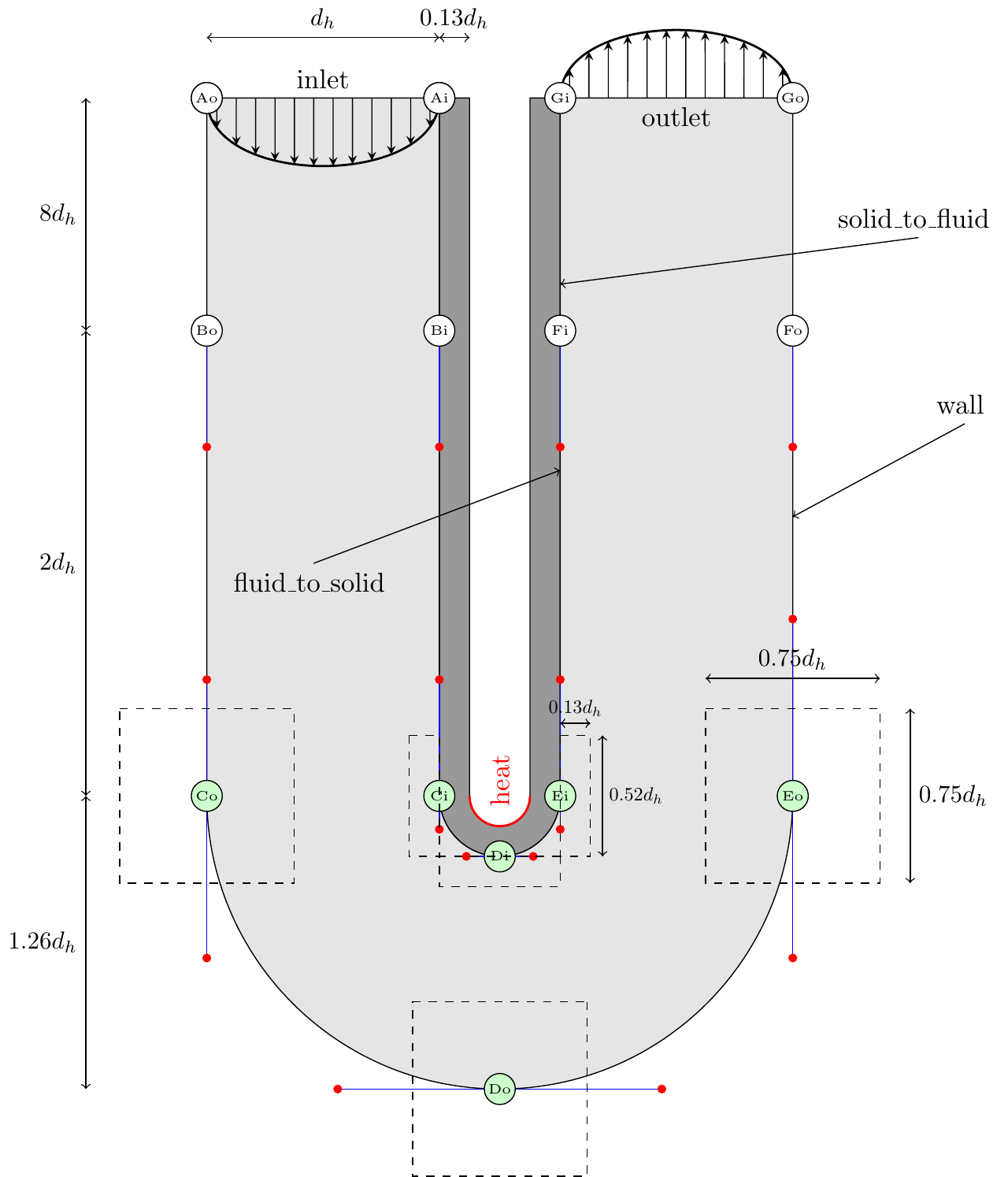}} 
	\caption{Parameterized initial geometry with boundary points and curve parameters. Modified from \cite{NDNET}}
    \label{fig:DesignSpaceParameter}
\end{figure}
 
 It has a circular U-bend with an outer radius of $1.26 \cdot d_h$, the fluid area is bound by an inner radius of $0.26 \cdot d_h$, and the solid with a thickness of $0.13 \cdot d_h$ is connected to it. The inlet and outlet sections have a length of $10 \cdot d_h$. The used hydraulic diameter is $d_h = 0.075m$. The initial U-bend with both regions, solid (dark grey) and fluid (light grey) is pictured in Figure~\ref{fig:DesignSpaceParameter}. The U-bend is partitioned into different levels A to G. These levels represent the position on a layer $o$ (outer) or $i$ (inner). Boundary points with white-filled circles are fixed and therefore provide no degrees of freedom to the entire system. However, boundary points with green-filled circles, represent two design parameters and thus contribute two degrees of freedom to the overall system. 
 In order not to display the complete length of the inlet and outlet channels the distance from level A/G to level B/F is depicted compressed in Figure~\ref{fig:DesignSpaceParameter}. Green-filled circles represent the boundary points of the geometry and can vary within the dashed boxes. These boxes represent the limits of the design parameters of the respective boundary point. The dimensions for the boxes on the outer layer are $0.75d_h$ x $0.75d_h$ and $0.52d_h$ x $0.13d_h$ on the inner layer. On the outer layer, each point with a green-filled circle is parameterized for both coordinates x and y between -1 and 1 and has its origin (initial solution) in the parameter values (0,0). Positive parameter values provide a widening, whereas negative values result in a constriction of the flow area. On the inner layer, one degree of freedom is also parameterized between -1 and 1. The other component is parameterized between 0~and~1. Third-order Bézier curves are used to connect the boundary points. Each curve is controlled by two curve design parameters which are represented in Figure~\ref{fig:DesignSpaceParameter} by the red dots. With the help of the blue lines the origin of each curve parameter is indicated. A red dot and its respective blue line can only move on a single axis (x or y). Each curve parameter value is between 0 and 1 and the value 0 places the curve parameter directly on the corresponding boundary point. Six boundary points and eight curves with two design parameters result in a total number of 28 design parameters each. In the following Figure \ref{fig:examples}, two example designs are presented.

\begin{figure}[ht]%
    \centering
    \subfigure{{\scalebox{1}{\includegraphics[width=.28\textwidth]{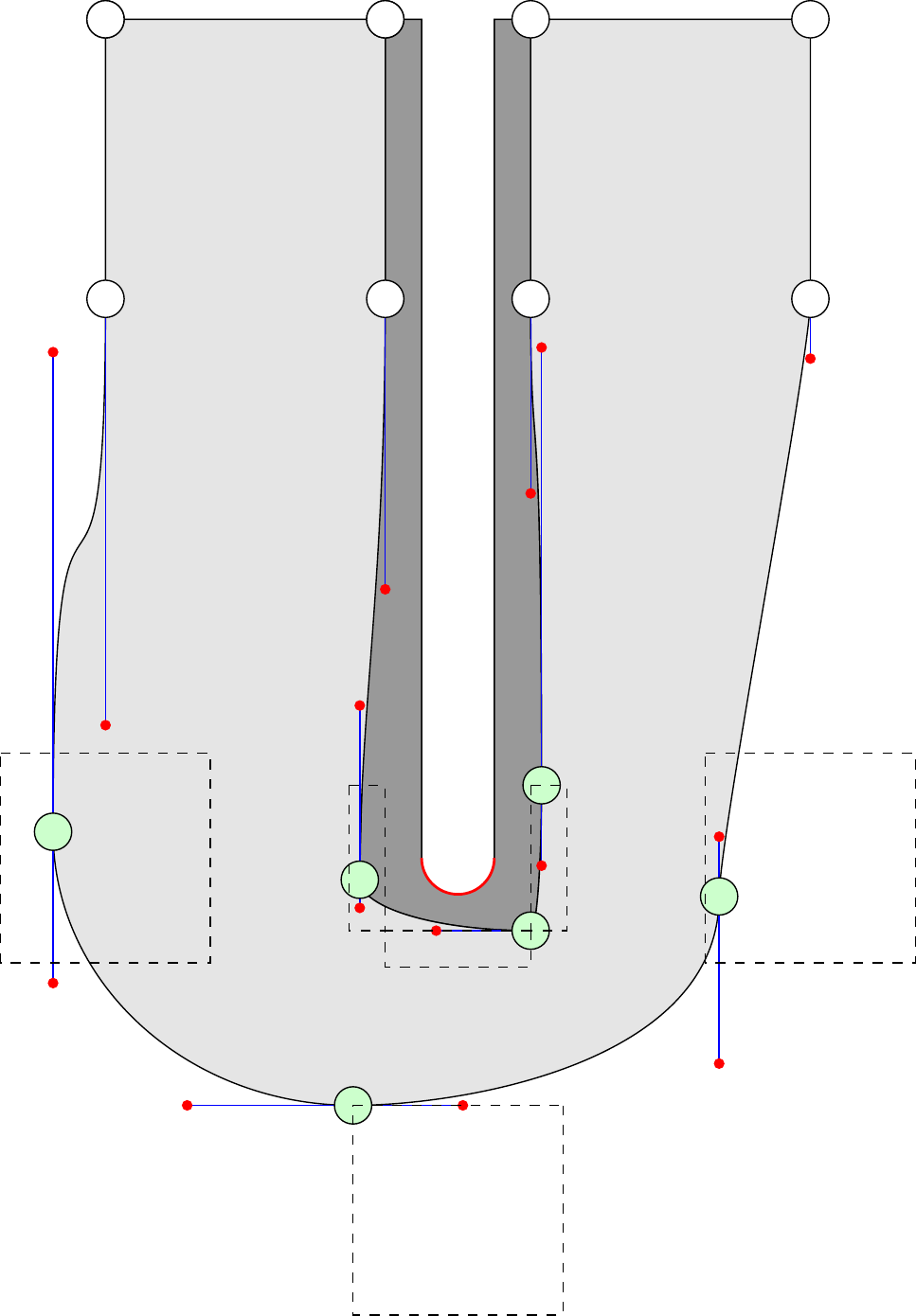} }}}%
    \qquad
    \subfigure{{\scalebox{1}{\includegraphics[width=.28\textwidth]{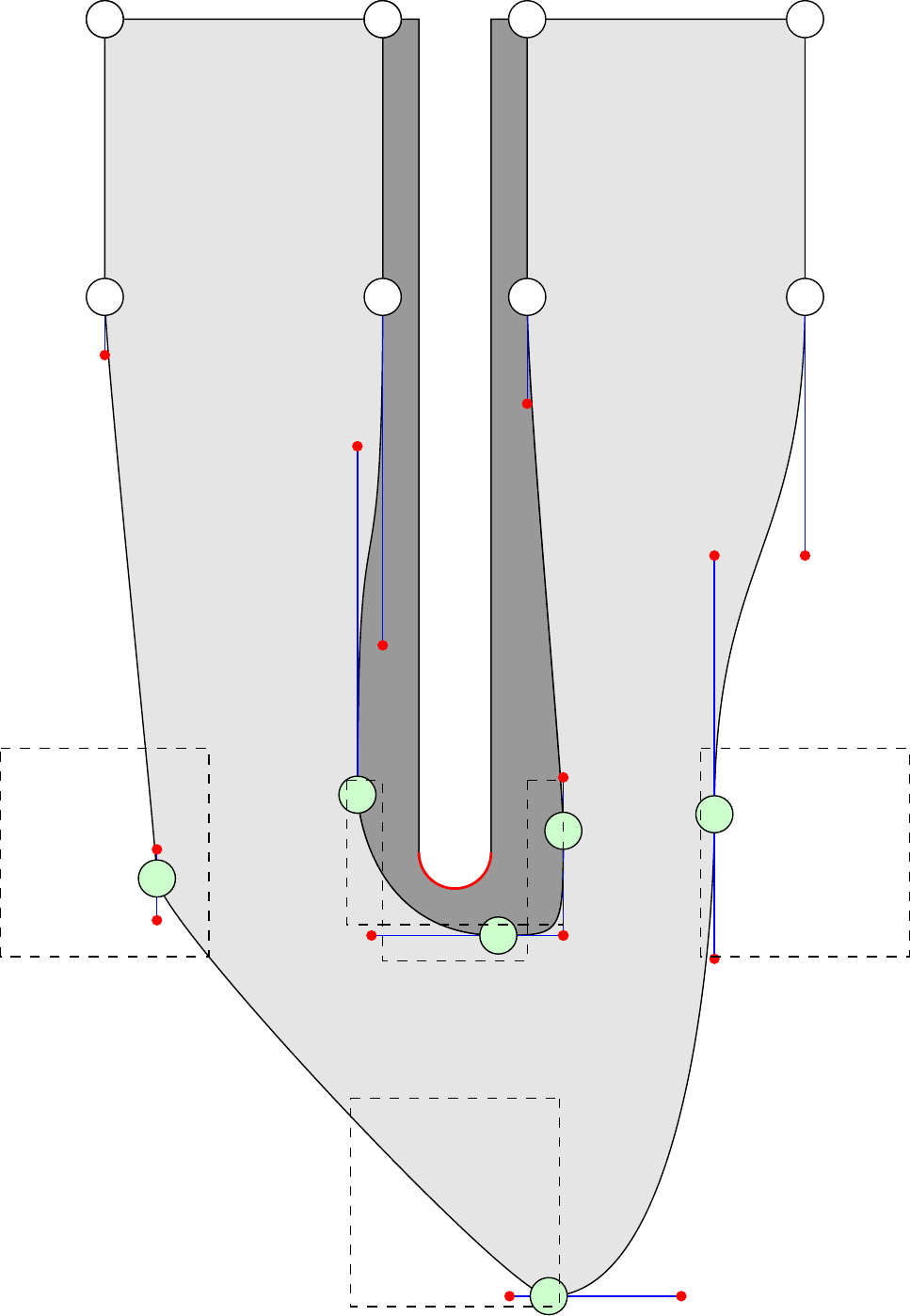}  }}}%
    \caption{Two samples with different design parameters}%
    \label{fig:examples}%
\end{figure}

\vskip0.1cm
\textbf{Material properties and boundary conditions:} In the following the basic material properties and used boundary conditions are presented. These are kept constant for each sample in the dataset. The U-bend depicted in Figure \ref{fig:DesignSpaceParameter} consists of a fluid and a solid region. The selected fluid is air and the solid region is made of construction steel. Material properties used are assumed to be constant and gathered in Table \ref{MD}.  
\begin{table}[!ht]
\caption{Material properties for the solid area (steel) and the fluid area (air)}
\label{MD}
\centering
\renewcommand{\arraystretch}{1.5}
\scriptsize
\begin{tabular}{c c c c c c} \hline
\textbf{material} & \textbf{thermal} \textbf{conductivity} & \textbf{heat capacity} & \textbf{density} & \textbf{Prandtl-number} & \textbf{kinematic viscosity} \\
\rowcolor{gray!25}
& $\lambda[W/(mK)]$ & $c_p[J/(kg K)]$ & $\rho[kg/m^3] $ & Pr[-] & $\nu[m^2/s]$ \\ \hline
solid & $25$ & $490$ & $7850$ & - & - \\ \hline
\rowcolor{gray!25}
fluid & $0.0262$ & $1004.5$ & $1.2$ &  $0.7$ & $1.5 \cdot 10^{-5}$ \\ \hline
\end{tabular}
\end{table}

Boundary conditions for each area are specified separately and described according to the names commonly used in OpenFOAM \cite{greenshields2022}. The \textit{noSlip} condition applies to the flow on all walls. This means that the flow velocity must be zero. A fixed value is specified at the \textit{inlet}. In normal direction, the gradient of the velocity must be zero at the \textit{outlet} due to the free flow. The fluid enters with a temperature of 300K. The pressure $p_{rgh}$ is the static pressure $p$ which is reduced by the hydrostatic pressure and is calculated at the entrance. On the adiabatic \textit{walls} and at the limit wall the \textit{fixedFluxPressure} boundary condition is used. At the outlet, the pressure $p_{rgh}$ is specified with a fixed value of zero. At each boundary the static pressure $p$ is calculated. At the surface \textit{heat}, a specific heat flux \textit{externalWallHeatFluxTemperature} of $q=100,000 W/m^2$ is given. In Figure \ref{fig:DesignSpaceParameter} all boundaries are shown and explained in detail for the fluid region in Table \ref{2D_RB_Fluid} and for the solid region in Table \ref{2D_RB_Solid}.

\begin{table}[!ht]
\caption{Boundary conditions of the fluid}
\centering
\footnotesize
\renewcommand{\arraystretch}{1.2}
\centering
\begin{tabular}{c c c c c} \hline
  & \textbf{inlet} & \textbf{outlet} & \textbf{fluid}$\_$\textbf{to}$\_$\textbf{solid} & \textbf{wall} \\ \hline
\rowcolor{gray!25}
Velocity $\textbf{v}$ & fixedValue & zeroGradient & fixedValue & noSlip \\
value & list of values & uniform (0 0 0) & uniform (0 0 0) & uniform (0 0 0) \\ \hline
\rowcolor{gray!25}
temperatur $T$ &  fixedValue & zeroGradient & turbulentTempreture- & zeroGradient \\
\rowcolor{gray!25}
 &  &  & CoupledBaffelMixed & \\
value & uniform 300 & uniform 300 & uniform 300 & uniform 0 \\ \hline
\rowcolor{gray!25}
pressure $p_{rgh}$ & fixedFluxPressure & fixedValue & fixedFluxPressure & zeroGradient \\
value & uniform 0 & uniform 0 & uniform 0 & uniform 0 \\ \hline
\rowcolor{gray!25}
pressure $p$ & calculated & calculated & calculated & calculated \\
value & uniform 0 & uniform 0 & uniform 0 & uniform 0  \\ \hline
\end{tabular}
\label{2D_RB_Fluid}
\end{table} 

At the interface between fluid and solid the temperature must be the same, therefore the heat flow supplied to the fluid must correspond to the heat conducted from the solid. With \textit{turbulentTemperatureCoupledBaffleMixed}, these two boundary conditions are realized and must be specified for both regions. Therefore, two boundaries instead of one boundary needs to be defined. For the fluid region, the limit is defined as \textit{fluid\_to\_solid}. It is the other way around for the solid region and is defined as \textit{solid\_to\_fluid}.  

\begin{table}[!ht]
\centering
\caption{Boundary conditions of the solid}
\begin{tabular}{c c c c c} \hline
  & \textbf{heat} & \textbf{wall} & \textbf{solid}$\_$\textbf{to}$\_$\textbf{fluid} \\ \hline
\rowcolor{gray!25}
temperatur $T$ &  externalWallHeat- & zeroGradient & turbulentTempreture- \\
\rowcolor{gray!25}
 &  FluxTemperature &  & CoupledBaffelMixed  \\
value & uniform 300 & uniform 300 & uniform 300 \\ \hline
\rowcolor{gray!25}
pressure $p$ & calculated & calculated & calculated \\
value & uniform 0 & uniform 0 & uniform 0 \\ \hline
\end{tabular}
\label{2D_RB_Solid}
\end{table}

A Mach-number of 0.05 allows to use an \textit{incompressible} assumption. A completely turbulent flow is indicated by the Reynolds number of 40000. It was decided to employ the two-equation $k-\omega~SST$ model in order to take turbulence into account. Used boundary conditions for the turbulent kinetic energy $k$, turbulent viskosity $\nu_t$ and specific turbulent dissipation rate $\omega$ are mentioned in Table \ref{table:turbulence}.

\begin{table}[!ht]
\caption{Turbulence boundary conditions}
\centering
\footnotesize
\renewcommand{\arraystretch}{1.2}
\centering
\begin{tabular}{c c c c c } \hline
  & \textbf{inlet} & \textbf{outlet} &  \textbf{wall} & \textbf{frontAndBack} \\ \hline
\rowcolor{gray!25}
turbulent kinetic energy $k$ & fixedValue & inletOutlet & kqRWallFunction & empty  \\
value & list of values & internalField & internalField &  \\ \hline
\rowcolor{gray!25}
turbulent viskosity $\nu_t$ &  fixedValue & calculated & nutkWallFunction & empty \\
value & list of values & internalField & internalField &  \\ \hline
\rowcolor{gray!25}
specific turbulent   & fixedValue & inletOutlet & omegaWallFunction & empty  \\
\rowcolor{gray!25}
dissipation rate $\omega$ &  & & &\\
value & list of values & internalField & internalField &   \\ \hline
\end{tabular}
\label{table:turbulence}
\end{table} 

The boundary condition \textit{inletOutlet} is usually the same as \textit{zeroGradient}. Nevertheless, it changes to a \textit{fixedValue} when the velocity vector targets into the domain next to the boundary. This \textit{fixed value} is the value at the \textit{inlet}.\\ 
In order to ensure a fully developed velocity profile at the \textit{inlet}, a one-dimensional channel flow is precomputed. Furthermore, turbulence values $k$, $\nu_t$ and $\omega$ are given from this precalculation. This ensures a suitable initial solution for the boundaries of the 2D U-Bend duct.\\
Using the initial geometry a comprehensive mesh study was performed. The most reliable results were obtained by using a mesh with 60 cells along the cross section in the fluid region, 15 in the solid region, and 780 cells along the flow course. The initial solution result was matched and validated with the help of measurements from Coletti et al. \cite{coletti_optimization_2013} using particle image velocimetry and a 3D simulation of Hayek et al. \cite{hayek_adjoint_based_2018}. To save computation time the converged numerical solution of the initial problem is used as a starting point for the later generation of samples.\\ 
\vskip0.1cm
\textbf{Objective functions:} Two objective functions are introduced to evaluate the different designs and to measure their performance. An obvious optimization problem in fluid mechanics is the minimization of pressure loss. The pressure loss
\begin{equation}
	J_1 = \dfrac{1}{A_{in}}\int_{A_{in}}p_{in}\,\mathrm{d}A_{in}-\dfrac{1}{A_{out}}\int_{A_{out}}p_{out}\,\mathrm{d}A_{out}
	\label{J1} 
\end{equation}
is calculated by the difference of the integral of the pressure $p$ over the area $A$ and normalized by the area $A$ of the inlet and outlet. A low pressure loss is accompanied by a low pumping effort. One important aspect is to keep the boundary condition for the volume flow, i.e. the quantity of pumped medium, constant, since a zero flow rate would result in a zero pressure loss. \\
The second objective function is used to quantify the cooling performance, which is done with the help of the temperature at the heating surface. A lower heating surface temperature with a defined heat flow results in a bigger heat transfer. Objective function 
\begin{equation}
	J_2 = \int_{A_{heat}}(T_{heat}-T_{in})^2\,\mathrm{d}A_{heat}
	\label{J2} 
\end{equation} 
is composed of the quadratic temperature difference between heating wall temperature $T_{heat}$ and inlet temperature $T_{in}$ integrated over the heating surface $A_{heat}$. This objective was used and introduced by Goeke~and~Wünsch in~\cite{goeke1}. From a physical perspective, the objectives are mutually exclusive, since no design minimizes both objectives. The reason for this is that convective heat transfer and pressure loss are coupled by Reynolds number, flow velocity and wall shear stress, respectively. $J_1$ would favor a lower Reynolds number. However, since a high Reynolds number benefits a high convective heat transfer, this would lower the objective $J_2$.
\vskip0.1cm
\textbf{Experimental design:} 
The workflow of the computer experiment is shown in Figure~\ref{fig:Experiment}. Only free and open-source software was used to set up the experiment. The programming language Python \cite{van1995python}, the workload manager Slurm~\cite{Yoo2003SLURMSL}, the container software Podman~\cite{podman} (to paravirtualize a Linux operating system) and OpenFOAM~\cite{greenshields2022} as a CFD software were applied. \\

\begin{figure}[ht]
	\centering
    \includegraphics[width=0.85\textwidth]{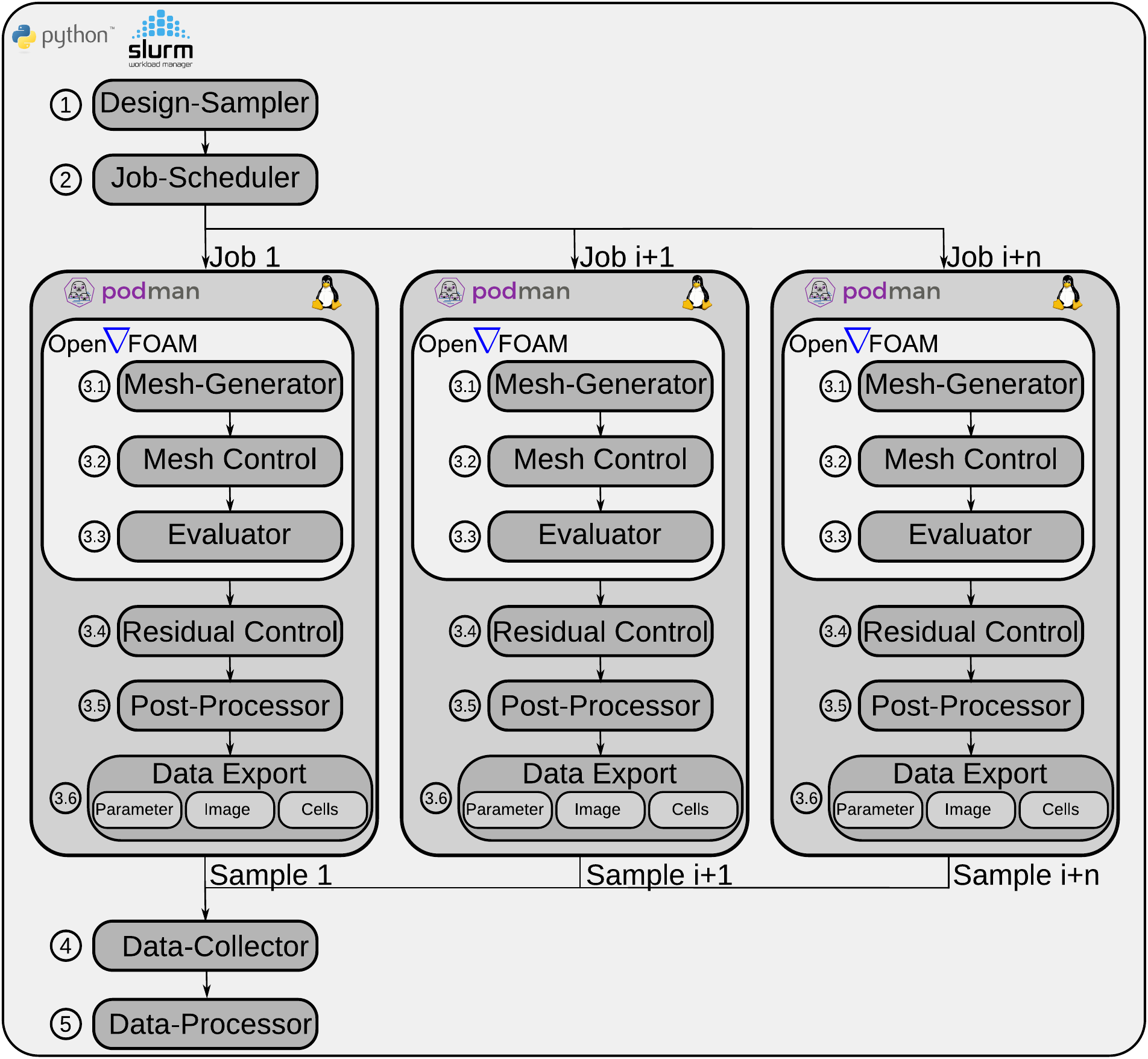}
	\caption{Experimental design to produce the U-bend design dataset}
    \label{fig:Experiment}
\end{figure}

The~~\textbf{\kreis{1}~Design-Sampler} generates a vector of 28 independent and i.i.d. random variables. In this step they are further processed into a format that can be read by OpenFOAM to create the mesh. Subsequently, the~~\textbf{\kreis{2}~Job-Scheduler} is used to parallelize a predefined number of jobs. The Job-Scheduler creates a job for each sample, that is fully automated and able to schedule jobs across multiple nodes in the computing cluster. Once simulations are completed, the results are gathered by the~~\textbf{\kreis{4}~Data-Collector}, so that new jobs are scheduled until the number of desired samples has been generated. \\
Inside the three large and parallel dark grey boxes the process of a simulation is shown in Figure~\ref{fig:Experiment}. With the help of the Podman paravirtualization software, a container is created for each simulation. This container runs a Linux-based guest operating system, the CFD software OpenFOAM and other auxiliary utilities that are required to process the data as desired. The file provided by the \textit{Design-Sampler} is received by the~~\textbf{\kreis{\tiny{3.1}}~Mesh-Generator} to generate the computational mesh. Now, the quality of the mesh is checked in~~\textbf{\kreis{\tiny{3.2}}~Mesh Control}. Since the solution quality of numerical simulations is very sensitive to the quality of the mesh a sample with a mesh that does not fulfill defined quality criteria is discarded. In addition, the parameterization allows the inner and outer layers to intersect, which is physically impossible and thereby prohibited. The parameters used to evaluate the mesh quality criterion are the maximal skewness, the maximal aspect ratio and the maximal non-orthogonality of single cells in the mesh.\\
The \textit{chtMultiRegionSimpleFoam} solver is used as~~\textbf{\kreis{\tiny{3.3}}~Evaluator}. It is a steady-state solver for buoyant, turbulent fluid flow and solid heat conduction with conjugate heat transfer between solid and fluid regions. The~~\textbf{\kreis{\tiny{3.4}}~Residual Control} monitors the quality of the solution, using the residuals of the individual numerical solution variables. Residuals are monitored for two reasons. Firstly, to stop the simulation as soon as a certain solution quality has been achieved and secondly, to filter samples that do not meet the minimum requirements for the solution quality after the simulation has been completed. The simulation of a single sample using an AMD EPYC 7002 processor (boost clock rate up to 3.35GHz) lasts between 15 and 240 minutes. Samples that do not have sufficient mesh quality or sufficient solution quality are penalized with artificially high objective values. Whether these samples should be considered for further investigations or excluded depends on the research question to be investigated and therefore they are basically still part of the dataset. The~~\textbf{\kreis{\tiny{3.5}}~Post-Processor} calculates the defined objective values from the numerical solution and passes them to the~~\textbf{\kreis{\tiny{3.6}}~Data Export}. There the solutions are prepared and exported according to Figure~\ref{fig:dataset} in the demonstrated data formats and forwarded to the~~\textbf{\kreis{4}~Data-Collector}. As a last step, the~~\textbf{\kreis{5}~Data-Processor} deletes unnecessary log files. A *.json file is compiled to represent the existing folder and data structure of the dataset. This *.json file allows the data to be used quickly and efficiently for Deep Learning tasks.    
\vskip0.5cm
\textbf{Ethics statements}\\
\noindent 
This study does not involve experiments on humans or animals.
\vskip0.3cm
\noindent
\textbf{CRediT author statement}\\
\noindent
\textbf{Jens Decke}: Conceptualization, Methodology, Software, Investigation, Data curation, Writing- Original draft, Visualization. \textbf{Olaf Wünsch}: Writing - Review \& Editing, Supervision, Resources.  \textbf{Bernhard Sick}: Resources, Project administration, Funding acquisition. \\
\noindent
\vskip0.3cm

\textbf{Acknowledgments}\\
The first author would like to express his sincere gratitude to Stephan Goeke for his unwavering support, meaningful discussions, and valuable suggestions throughout the project. His exceptional collaboration and willingness to share ideas have been instrumental in shaping this research.

Special thanks are expressed to the server team of the Intelligent Embedded Systems department, in particular to Dr.~Christian~Gruhl and David~Meier, who customized some features on the compute cluster for the purpose of creating this dataset.\\  

This research has been funded by the Federal Ministry for Economic Affairs and Climate Action (BMWK) within the project "KI-basierte Topologieoptimierung elektrischer Maschinen (KITE)" (19I21034C). 
\vskip0.3cm
\textbf{Declaration of Competing Interest}\\
The authors declare that they have no known competing financial interests or personal relationships that could have appeared to influence the work reported in this paper.
\vskip0.3cm

\bibliographystyle{ieeetr}
\bibliography{refs}

\end{flushleft}
\end{document}